# Normal Metal Hot-Electron Nanobolometer with Johnson Noise Thermometry Readout

Boris S. Karasik, *Member IEEE,* Christopher B. McKitterick, Theodore J. Reck, and Daniel E. Prober

*Abstract*—The sensitivity of a THz hot-electron nanobolometer (nano-HEB) made from a normal metal is analyzed. Johnson Noise Thermometry (JNT) is employed as a readout technique. In contrast to its superconducting TES counterpart, the normal-metal nano-HEB can operate at any cryogenic temperature depending on the required radiation background limited Noise Equivalent Power (NEP). It does not require bias lines; 100s of nano-HEBs can be read by a single low-noise X-band amplifier via a filter bank channelizer. The modeling predicts that even with the sensitivity penalty due to the amplifier noise, an NEP $\sim 10^{-20} - 10^{-19}$ W/Hz$^{1/2}$ can be expected at 50-100 mK in 10-20 nm thin titanium (Ti) normal metal HEBs with niobium (Nb) contacts. This NEP is fairly constant over a range of readout frequencies $\sim 10$ GHz. Although materials with weaker electron-phonon coupling (bismuth, graphene) do not improve the minimum achievable NEP, they can be considered if a larger than 10 GHz readout bandwidth is required.

## I. INTRODUCTION

Recently we analyzed the performance of the graphene HEB detector using JNT as a readout [1, 2]. The motivation for that was a desire to increase the saturation power and simplify the array architecture of sensitive far-IR/THz astrophysics detectors. An attractive feature of the graphene HEB is a very low electron-phonon (e-ph) thermal conductance due to the extremely small volume of a device one monoatomic layer thick. However, the realization of such a detector requires more studies of the doping and disorder effects on the e-ph interaction in the material.

Our recent work focused on the hot-electron Transition-Edge Sensor (TES) where a much lower thermal conductance than in a SiN membrane suspended TES has been achieved [3]. This is due to the weak electron-phonon (e-ph) coupling in a micron- or submicron-size hot-electron Ti TES. Using this approach, the targeted low NEP $\sim 10^{-19}$ W/Hz$^{1/2}$ has been confirmed via direct optical measurements [4]. The kinetic inductance detector [5] and quantum capacitance detector [6] have demonstrated a similar sensitivity as well.

Even though the e-ph constant in graphene [7] may be very small compared to that in a conventional disordered normal metals (e.g., Ti [3, 8]) there are several practical considerations that make the latter attractive. They include the reliability of fabrication of many identical devices on the same wafer, the feasibility of a 25-100 Ω sheet resistance in the film needed for achieving a good impedance match between a 1-2 square long device and a planar antenna, small or no contact resistance between the device and superconducting Andreev contacts made from Nb, NbN, or NbTiN.

In this paper, we analyze the expected sensitivity (NEP) of a Ti hot-electron nanobolometer embedded into a planar antenna or waveguide circuit via superconducting contacts with critical temperature $T_C \sim 10$ K. The well-established data on the strength of the e-ph coupling are used in the present analysis and the contribution of the readout noise into the NEP is explicitly computed. The readout scheme utilizes the JNT allowing for Frequency-Domain Multiplexing (FDM) using coupling of HEBs via a narrowband filter-bank channelizer. Such filters may be implemented using coupled superconducting microresonators which, in a way, are similar to many current FDM schemes for reading TES and kinetic inductance detectors (KID) [9]. The room temperature electronics for the JNT FDM will a digital spectrometer rather than a more complex phase sensitive transceiver ultilized with other detectors (see, e.g., [10]).

It turns out that the filter bandwidth and the summing amplifier noise define the overall system sensitivity. The analysis shows that the noise bandwidth (the readout frequency range within which NEP is constant) of the Ti HEB is about 7-10 GHz even at temperatures of 50-100 mK. Materials with weaker e-ph coupling (Bi, graphene) may have even larger noise bandwidth. Beside the high sensitivity, the normal metal bolometer does not have any hard saturation limit and thus can be used for imaging with arbitrary contrast. By changing the operating temperature the bolometer sensitivity can be fine tuned to the background noise in a particular application. Using a broadband quantum noise limited kinetic inductance parametric amplifier [11], 100s of normal metal HEBs can be read simultaneously without saturation of the system output.

## II. TITANIUM NANO-HEB

Titanium superconducting nano-HEBs have been studied by us over the past several years (see [3] for overview). The devices were Transition-Edge Sensors (TES) so proper voltage bias with a SQUID amplifier was required for readout without adding noise. The device noise was limited by the fundamental thermal energy fluctuations (TEF), "phonon" noise. Submicron size (down to 1μm × 0.13μm × 40nm) superconducting devices have been achieved with a very low e-ph thermal conductance $G_{e-ph}$ = 0.3 fW/K at 65 mK [12]. In order to achieve such a low thermal conductance, superconducting contacts made from Nb or NbTiN were required. The former were achieved using evaporation of Ti and Nb in vacuum through a shadow mask [12] whereas the latter were done using magnetron sputtering [4].

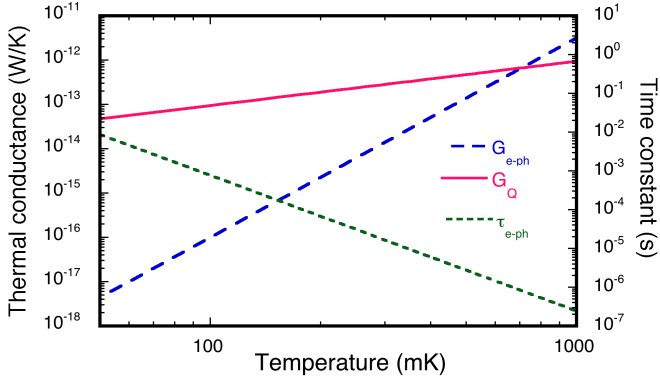

Fig. 1. Electron-phonon thermal conductance in a Ti normal metal HEB with dimensions 0.5μm × 0.25μm × 20nm, "quantum of thermal conductance," and electron-phonon time constant as functions of temperature.

During the course of our work we found empirically that small thin-film Ti devices with Nb contacts often ceases to be superconductive when prepared via magnetron sputtering deposition in a chamber where plasma etch has been used. In contrast, similarly fabricated Ti devices with NbTiN contacts were superconductive. The apparent cause for the disappearance of superconductivity in Ti has something to do with either the electrochemical reaction between the two metals in the presence of residual Cl, F, and H$_2$O, or with the photochemical reaction in acetone during the resist removal though the exact mechanism has never been fully understood. At the same time, the Nb contacts remained superconductive. The normal state resistivity of the bulk of Ti devices was ρ = 30-40 μΩ cm thus it is fairly straight forward to achieve a 50-Ω normal resistance in 1-2 square devices made from a 10-20 nm thick film.

### III. THERMAL CONDUCTANCE

The electron-phonon thermal conductance in the normal metal HEB is nearly the same as in a TES HEB so we can apply the previously obtained data for the device proposed here. Figure 1 demonstrates the expected e-ph thermal conductance $G_{e-ph}$ and the corresponding time constant, $\tau_{e-ph}$, for a hypothetical Ti device with dimensions 0.5μm × 0.25μm × 20nm which can be achieved using a UV photolithography on sapphire or Si$_x$N$_y$. These two substrates provide the lowest e-ph thermal conductance (a discussion of the substrate effect is given in [3]) which can be expressed as $G_{e-ph}(T) = m\Sigma V T^{m-1}$ where $V$ is the device volume, $m$ = 5.5, $\Sigma$ = 2.3 × 10$^8$ W/(K$^{4.5}$ m$^3$) [3]. The e-ph thermal conductance in a nano-HEB is very low below 100 mK. Preventing thermal energy from leaking away, thus bypassing the e-ph channel is a critical requirement. Good quality superconducting contacts with a large energy gap dramatically reduce the heat diffusion to the contacts, due to Andreev reflection [13]. In contrast, with non-superconducting contacts, fast outdiffusion of heat to the contacts would occur, giving strong diffusion cooling. The characteristic time for that outdiffusion cooling is $\tau_{diff} = L^2/(\pi^2 D)$ [14, 15] where $L$ is the bridge length and $D$ is the electron diffusion constant ($D$ ≈ 2 cm$^2$/s for our Ti films [8]). For $L$ = 0.5 μm, $\tau_{diff}$ = 130 ps. However, in the presence of superconducting contacts with large gap $\Delta$ only a minuscule fraction of electrons ξ ≈ 1.44•exp(-$\Delta/k_B T$) can contribute to the heat diffusion [1]. In Nb, $\Delta$ ≈ 1.5 meV therefore the effective cooling rate per electron due to diffusion is $(\tau_{diff}^*)^{-1}$ = ξ/$\tau_{diff}$ ~ 10$^{-141}$ s$^{-1}$ at 50 mK whereas the cooling rate due to phonons is 1/$\tau_{e-ph}$ ~ 10$^2$ s$^{-1}$ [8].

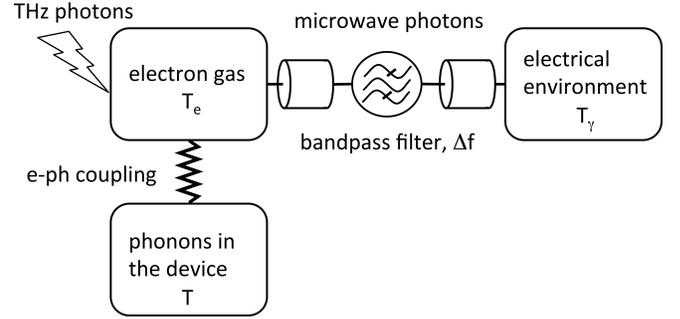

Fig. 2. Cooling pathways for hot electrons in a normal metal HEB. As in TES HEBs, electrons cool via emission of acoustic phonons in the device. Also, because of the necessity to read the Johnson noise within a bandwidth $\Delta f$, additional cooling occurs via emission of microwave phonons.

A much more significant contribution to the cooling comes from the emission of microwave radiation via connecting wires. Figure 1 shows the total thermal conductance (aka "quantum of thermal conductance") $G_Q = \pi^2 k_B^2 T/3h$ ≈ 1 pW/K × $T$ due to the microwave emission in a single mode [16]. This includes all the photons emitted by a 1D blackbody over the frequency range $\Delta f$ ~ $k_B T/h$ ~ GHz. In the case of the TES HEB, it was relatively easy to minimize this thermal conductance by using a lowpass cold filter leaving only enough bandwidth $\Delta f$ ~ 100 kHz for reading the HEB with a SQUID. In this case, the microwave photon mediated thermal conductance $G_\gamma = k_B \Delta f$ = 1.4×10$^{-18}$ W/K which is smaller than $G_{e-ph}$ even at 50 mK.

For the normal metal HEB, Johnson noise is used to read the electron temperature $T_e$ directly so some bandwidth of the order of MHz centered at GHz frequency is required (see Fig. 2). The need for GHz readout is dictated by the desire to frequency multiplex many devices with a single broadband amplifier. Each of the pixels is connected to a filter tuned to an individual frequency. In this case, the heat will be carried away by means of emission of microwave photons. The

---
[1] A normalization prefactor of 1.44 is due to the Fermi distribution of electrons.

corresponding thermal conductance is [17]:

$$G_\gamma(T,f) = hf\, \partial n(T,f)/\partial T \Delta f,  \quad (1)$$

where $n(T,f) = [\exp(hf/k_BT)-1]^{-1}$ is the photon occupation number. In the case $n \gg 1$ (Rayleigh-Jeans limit), $n \approx k_BT/hf$ and $G_\gamma \approx k_B \Delta f$. In the opposite case $n \approx \exp(-hf/k_BT) \ll 1$ (Wien's limit):

$$G_\gamma \approx k_B \Delta f (hf/k_BT)^2 n, \quad (2)$$

that is, $G_\gamma$ becomes very small.

Figure 3 illustrates the variation of $G_\gamma$ with temperature and frequency for a fixed bandwidth of $\Delta f = 10$ MHz. It decreases significantly by ~ 10 GHz at low temperature 50-100 mK.

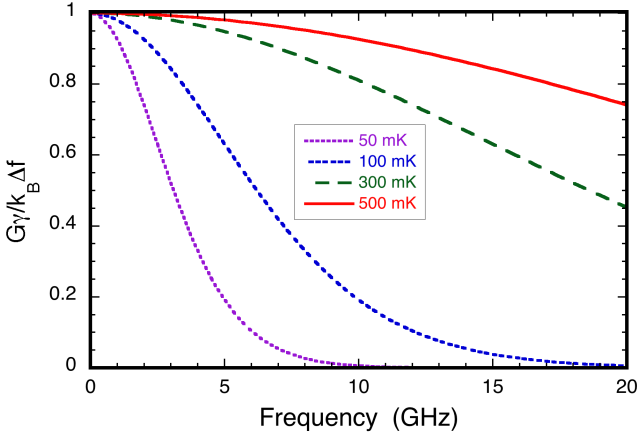

Fig. 3. Microwave photon mediated thermal conductance as function of temperature and frequency ($\Delta f = 10$ MHz).

## IV. JOHNSON NOISE READOUT AND NEP

The fundamental TEF noise limits the HEB NEP according to the following expression:

$$NEP_{TEF} = \sqrt{4k_BT^2(G_{e-ph}+G_\gamma)}. \quad (3)$$

However, achieving this limit in the normal metal HEB is harder than in, e.g., TES HEB. In the latter, the responsivity to radiation power $P_{rad}$ (e.g., current responsivity $dI/dP_{rad}$) is very large. This makes the contribution of the readout (SQUID) noise very small or often negligible. In the normal metal HEB, microwave power emitted by the bolometer $P_f = hf\Delta f n$ plays the role of the output signal. Thus the responsivity is $dP_f/dP_{rad}$.[2] The output noise is a sum of the Johnson noise and the amplifier noise. The power of the Johnson noise is a variance of $P_f$ [18] and can be expressed as follows:

$$P_J = hf\Delta f \sqrt{n(n+1)}. \quad (4)$$

(Note, for $n \gg 1$, $P_J$ reduces to a classic expression $P_J = k_BT\Delta f$). Due to the narrowband nature of the readout, an effective noise temperature at the output can be introduced: $T^* = P_J/k_B\Delta f$. Together with the amplifier noise temperature, $T_A$, it determines the effective temperature fluctuation due to the

---

[2] This definition of the responsivity is a corrected form of Eq.13 in [1].

system noise (Dicke formula) [19]:

$$\delta T = (T^* + T_A)/\sqrt{t_a \Delta f}. \quad (5)$$

Here $t_a$ is the averaging time. Noting that an electron temperature increase in the small signal approximation is $\Delta T_e = P_{rad}/(G_{e-ph}+G_\gamma)$ and choosing $t_a = 0.5$ s (a 1-Hz output detection bandwidth) we can arrive to an expression for NEP due to the readout noise:

$$NEP_{JNT} = k_B \delta T \Delta f (G_{e-ph}+G_\gamma)(\partial P_f/\partial T_e)^{-1}. \quad (6)$$

After some transformations, this expression can be written as follows:

$$NEP_{JNT} = \frac{(k_BT/hf)^2 (G_{e-ph}+G_\gamma)\left[T_A+(hf/k_B)\sqrt{n(n+1)}\right]}{n(n+1)\sqrt{\Delta f/2}}. \quad (7)$$

Modeling results for NEP given by Eq. 3 and 7 are presented in Fig. 4. For the modeling, we assume $T_A = 0.5$ K which is the quantum limit at 10 GHz and $\Delta f = 10$ MHz. The data for 50 mK bear the most pronounced features of all the thermal and radiation effects discussed above. The TEF NEP decreases with frequency that corresponds to a gradual extinction of $G_\gamma$ (see Fig. 3 for comparison). At the same time, the total NEP (dominated by $NEP_{JNT}$) remains constant. This behavior can be easily understood due to the fact that for small values of $n$ ($n = 0.001$ at 7 GHz) $T^* \ll T_A$. Therefore,

$$NEP_{JNT} \approx k_BT_A\sqrt{2\Delta f} = 3\times 10^{-20}\, W/Hz^{1/2}. \quad (8)$$

This limit represents the fluctuation of the noise power of single mode thermal radiation with an effective temperature $T_A$.

At higher temperatures, the difference between $G_{e-ph}$ and $G_\gamma$ decreases. The frequency dependence vanishes when $G_{e-ph}$ eventually dominates. Other materials with potentially smaller value of $G_{e-ph}$ (e.g., graphene, bismuth) would exhibit a larger crossover frequency between $G_\gamma$ and $G_{e-ph}$. Thus a broader frequency range where $NEP_{JNT}$ is constant and given by Eq. 8 can be reached. However, the readout noise remains the largest component of the total NEP.

## V. READOUT AMPLIFIER AND MULTIPLEXING CONSIDERATIONS

Large bandwidth and low amplifier noise are the key requirements for the implementation of the sensitive normal metal HEB array. $T_A \sim 0.5$ K, which is used on this paper, was reported for, e.g., SQUID rf amplifiers at somewhat lower frequencies [20, 21]. A novel parametric amplifier concept utilizing non-linearity of the kinetic inductance in NbTiN [11] has demonstrated multi-GHz bandwidth and has achieved the noise temperature $T_A \approx 1.5$ K (3 photons) at 10 GHz [22]. The amplifier has a gain bandwidth ~ 6 GHz with a minimum gain of more than 10 dB. In this way, it will be noise matched with a second-stage cryogenic HEMT amplifier with $T_A \approx 5$ K.

In order to read many HEB pixels with a single amplifier the frequency domain multiplexer design utilizing a filter-bank channelizer is needed. A preliminary design for $f \approx 10$

GHz, $\Delta f$ = 10 MHz is already available. To achieve the narrow bandwidths required in a compact, easily arrayed structure, the straightforward topology shown in Fig. 5 can be used [23]. The channelizer consists of a microstrip resonator coupled to the detector and the main readout line with parallel coupled lines. The channel frequency is controlled by the length of the resonator and the bandwidth of the channel is set by the spacing of the gaps between the coupled lines. To avoid interaction between the channels, they are each separated by a quarter-wavelength. This allows each channel to be designed separately avoiding the complex optimization problem of other (directional and elliptical filters) channelizer designs [24, 25]. The simple structure of each filter allows for rapid optimization of the coupler spacing and transmission line widths, which significantly reduces the computational cost of a highly multiplexed channelizer.

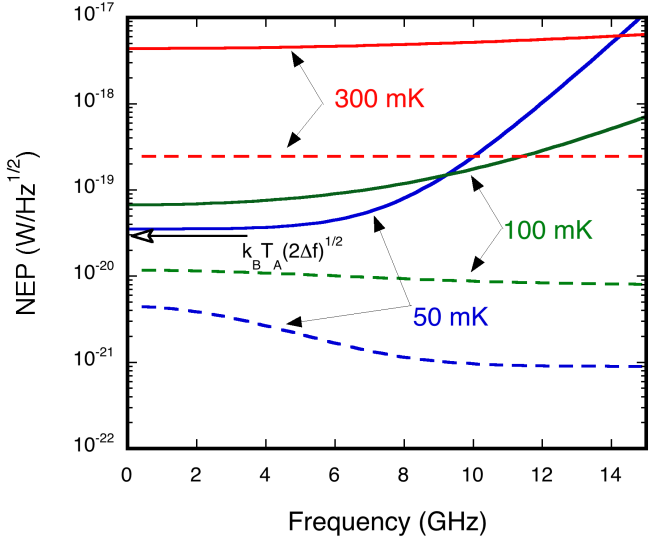

Fig. 4. Modeling results for $NEP_{TEF}$ (dashes) and $NEP_{JNT}$ due to the readout noise (solid lines).

The primary drawback to this approach is that when the filter is well matched to the multiplexed feed-line, only 50% of the energy is coupled to the amplifier. A more traditional approach to the channelizer design could be applied to improve the efficiency of the multiplexer, but the high-order filters required to achieve the 10 MHz channel bandwidth (0.1% at 10GHz) would make optimization of the full multiplexer challenging.

A simulation of a channel is shown in Fig. 5. The design is of a superconducting microstrip resonator on 300-μm thick silicon. Simulation of the channel's input coupling shows peak coupling of -3dB to the multiplexer readout and a 10 MHz -10dB bandwidth. The size of the filter is dominated by the length of the resonator, which could be reduced, if required by the pixel spacing, with a meandered transmission line or loop resonator [26].

In general, microresonators may exhibit a substantial noise due to the surface layer of two-level system fluctuators in the dielectric (see [9] for overview). It has been shown that this noise affects only the phase of a resonator and does not affect the amplitude. It has been shown [27] that the amplitude (dissipation) noise is at the vacuum noise level for a coplanar waveguide resonator made on high-purity intrinsic silicon. Since the JNT readout relies on the microwave noise power measurement the added noise from the resonator-based filter is likely to be very small. Use of crystalline silicon dielectric is also advanatageous compared to amorphous dielectric materials (e.g., $SiO_2$, α-Si:H, $Si_xN_y$) in view of its very low loss tangent, $\sim 10^{-5}$ [28]. Such a low loss does not affect the filter design of Fig. 5.

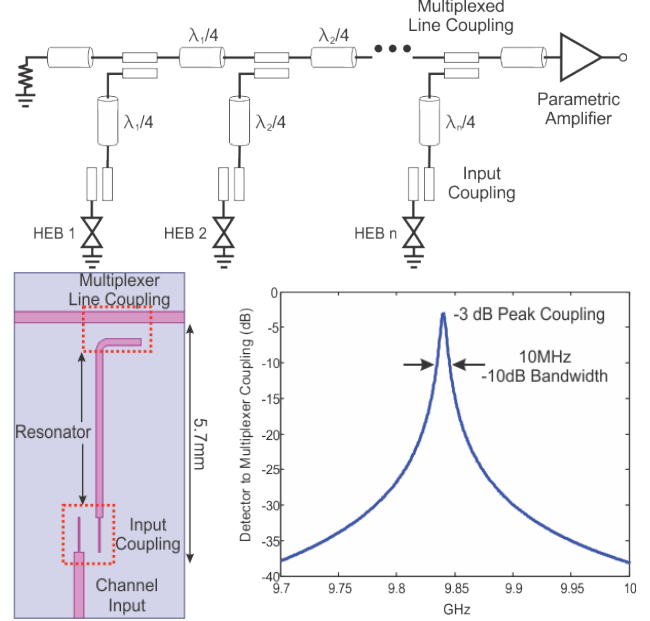

Fig. 5. Top: Schematic of a multiplexor using λ/4 resonators. Bottom left: Detailed view of the coupling geometry between the filter and the transmission line. Bottom right: Modeled coupling efficiency between a 50-Ω HEB element and an amplifier at 9.85 GHz.

Beside the presented 10-MHz filter design allowing for ~ 100 pixels an amplifier (~ 6 GHz amplifier bandwidth [11]), an additional analysis concerning the feasibility of a 1-MHz filter will be useful. Indeed, according to Eq. 8 a reduction of NEP can be achieved with a more narrowband filter. Moreover, this will allow for a 10-fold increase (to 1000s) of the number of pixels per one amplifier.

## VI. MAXIMUM OPTICAL LOAD AND DYNAMIC RANGE

By employing a normal (non-superconducting) absorber, we allow a significant increase of the electron temperature in the normal metal HEB without hard saturation of the output signal. A superconducting TES would show such hard saturation. The practical limit is set by the critical temperature of superconducting Andreev contacts (9.3 K for Nb). The radiation power causing such an increase of the electron temperature can be estimated from the heat balance equation:

$$\Sigma V\left(T_e^m - T^m\right) + hf\left[n(T_e) - n(T_\gamma)\right]\Delta f + \\ + \int_T^{T_e} C_e(T) dT / \tau_{diff}^*(T) = P_{rad} \quad (7)$$

Here the first term on the left part of equation is the heat exchange between electrons and phonons, the second term is the microwave power exchange between electrons and the amplifier input (electrical environment in Fig. 2), and the third term is the heat flux due to the electron diffusion. $C_e = \gamma V T$ is the electron heat capacity ($\gamma = 310$ J K$^{-2}$ m$^{-3}$ for Ti). For simplicity, we ignore the fact that exponent $m$ at 10 K may be different from that at 50 mK (see [29] for details). As discussed above, electron diffusion cooling is negligible at 50 mK due to the Andreev reflection. Its contribution is still minor (~ 4%) even when the electron temperature approaches $T_C$ in Nb contacts. The microwave power exchange plays a role only below 100 mK.

It follows from Eq. 7 that in order to reach $T_e \approx 10$ K starting from 50 mK an optical load of $P_{rad} \approx 7$ nW per pixel is required. This is a huge number in comparison with that for TES detectors with similar sensitivity. Even with $t_a = 10$ ms, the dynamic range in this case is $P_{rad}\sqrt{t_a}/NEP > 100$ dB. The parametric amplifier will not limit the output signal even for a 1000-pixel HEB array since it has a very large 1-dB gain compression input power level of -52 dBm (6.3 µW) [11].

## VII. Conclusion

The presented analysis demonstrates the potential of the normal metals nano-HEB for astrophysics THz arrays. Even though the NEP of the detector is readout limited it can be below $10^{-19}$ W/Hz$^{1/2}$ which is a targeted figure for future spectroscopic applications on space telescopes. The unique feature of this detector is its very high dynamic range (no hard saturation of the detector output) and possibility to work at any cryogenic temperature depending on the expected background optical load. A relative simplicity of the normal metal HEB compared to the TES version (no need in bias, tuning of critical temperature, or dc SQUID amplifiers) and its large multiplexing ratio (100s pixels per single amplifier) can make this approach attractive for various space borne, airborne, and ground based instruments.

## Acknowledgment

BSK would like to thank A. Sergeev and P. Day for numerous discussions and R. Cantor for the previous collaborative work leading to realization of the feasibility of Ti normal metal nano-HEB.